\documentclass[11pt,fleqn]{article}

\usepackage{amsmath,amssymb,amsthm,enumerate}

\newcommand{\p}{\partial}
\newcommand{\const}{\mathop{\rm const}\nolimits}

\newcommand{\sign}{\mathop{\rm sign}\nolimits}

\newcounter{tbn}

\newcounter{mcasenum}

\newtheorem*{proposition*}{Proposition}
{\theoremstyle{definition}

}

\begin{document}

\par\noindent {\LARGE\bf
Exact Solutions of a Remarkable Fin Equation
\par}

{\vspace{4mm}\par\noindent {\it R.~O.~Popovych$^{\dag\ddag 1}$, C.~Sophocleous$^{\S 2}$ and O.~O.~Vaneeva$^{\dag 3}$}
\par\vspace{2mm}\par} {\vspace{2mm}\par\noindent {\it
${}^\dag$Institute of Mathematics of NAS of Ukraine, 3 Tereshchenkivska Str., Kyiv-4, 01601 Ukraine\\
$^\ddag$Fakult\"at f\"ur Mathematik, Universit\"at Wien, Nordbergstra{\ss}e 15, A-1090 Wien, Austria\\
$^\S$Department of Mathematics and Statistics, University of Cyprus, Nicosia CY 1678, Cyprus\\
}} 
{\noindent {\it $^1$rop@imath.kiev.ua, $^2$christod@ucy.ac.cy, $^3$vaneeva@imath.kiev.ua}\par}

{\vspace{5mm}\par\noindent\hspace*{8mm}\parbox{146mm}{\small 
A model `remarkable' fin equation is singled out from a class of nonlinear $(1+1)$-dimensional fin equations. 
For this equation a number of exact solutions are constructed by means of using both classical Lie algorithm and 
different modern techniques 
(functional separation of variables, generalized conditional symmetries, hidden symmetries etc). 
}\par}

\section{Introduction}

Mathematical models of conductivity and diffusion processes are traditional objects for investigations with  
symmetry methods and related to them approaches~\cite{Ibragimov1994V1}. 
Since these models are often formulated in terms of nonlinear differential equations which are, as rule, non-integrable and 
cannot be linearized, symmetry methods are important for construction of their exact solutions. 
It is indicative that modern development of group analysis of differential equations was begun from 
the study of a class of $(1+1)$-dimensional nonlinear diffusion equations~\cite{Ovsiannikov1959}. 
Afterward a wide range of diffusion equations was investigated within the symmetry framework 
(see e.g.\ \cite{Basarab-Horwath&Lahno&Zhdanov2001,Dorodnitsyn1982,Galaktionov1990,Ibragimov1994V1,
Lahno&Spichak&Stognii2002,Vaneeva&Johnpillai&Popovych&Sophocleous2006}). 
At the same time, even `simplified' $(1+1)$-dimensional non-linear diffusion models are fraught with a great many of `symmetry mysteries' 
which remain to be solved.  

Recently the class of nonlinear fin equations of the general form 
\begin{equation}\label{EqFin}
u_t=(D(u)u_x)_x+h(x)u,
\end{equation}
was investigated with the symmetry point of view in a number of 
papers~\cite{Bokhari&Kara&Zaman2006,Pakdemirli&Sahin2004,Pakdemirli&Sahin2006,Vaneeva&Johnpillai&Popovych&Sophocleous2006b}. 
Here $u$ is treated as the dimensionless temperature, 
$t$ and $x$ the dimensionless time and space variables, 
$D$ the thermal conductivity, $h=-N^2f(x)$, $N$ the fin parameter and $f$ the heat transfer coefficient.
(See e.g.~\cite{Bokhari&Kara&Zaman2006} for references on physical meaning and applications of equations~\eqref{EqFin}.)

Note that certain partial equations from class~\eqref{EqFin} were studied formerly.  
For example, the condition $D_u=0$ corresponds to the linear case of~\eqref{EqFin} 
which was completely investigated with the Lie symmetry point of view long time ago~\cite{Lie1881,Ovsiannikov1982}. 
The problem of group classification for the class of nonlinear one-dimensional diffusion equations 
(the degenerate case $h=0$) was first solved by Ovsiannikov~\cite{Ovsiannikov1959,Ovsiannikov1982}. 
The class of diffusion--reaction equations classified by Dorodnitsyn~\cite{Dorodnitsyn1982,Ibragimov1994V1} 
contains the equations of form~\eqref{EqFin} with $h=\const$. 
Group classification of the subclass where the thermal conductivity is a power function of the temperature was 
carried out in~\cite{Vaneeva&Johnpillai&Popovych&Sophocleous2006}. 
Large sets of exact solutions were constructed for the above equations and collected e.g.\ in~\cite{Ibragimov1994V1,Polyanin&Zaitsev2004}. 

In contrast to~\cite{Bokhari&Kara&Zaman2006,Pakdemirli&Sahin2004,Pakdemirli&Sahin2006}, 
the study in~\cite{Vaneeva&Johnpillai&Popovych&Sophocleous2006b} was concentrated on rigorous and exhaustive group classification 
of the whole class~\eqref{EqFin} and on construction of exact solutions for truly nonlinear and 'variable-coefficient' 
equations from this class. To find exact solutions, both classical Lie reduction and different modern approaches were applied.  
Although some interesting exact solutions were constructed, almost all of them are either stationary or scale-invariant. 
Therefore, the problem on more complicated exact solutions of equations~\eqref{EqFin} with $D_uh_x\not=0$ remains still open
even for partial values of the parameter-functions. 

In this letter we single out a model `remarkable' fin equation~\eqref{EqFin} 
with the fixed values of the parameter-functions $D=u^{-3/2}$ and $h=x^{-1}$,~i.e. 
\begin{equation}\label{EqFinRemarkable}
u_t=(|u|^{-3/2}u_x)_x+x^{-1}u,
\end{equation}
and investigate it in detail. 
A number of its exact solutions are constructed with a variety of symmetry techniques 
(Lie reduction, non-linear separation of variables, generalized conditional symmetries, hidden symmetries etc) 
in a closed form.

\section{Lie invariance}

Equation~\eqref{EqFinRemarkable} is ordinary, from the Lie symmetry point of view, in class~\eqref{EqFin}. 
The maximal Lie invariance group of~\eqref{EqFinRemarkable} is $A_1=\langle\p_t,\ D=t\p_t+x\p_x-\tfrac23u\p_u\rangle$, 
i.e. its Lie symmetry group~$G_1$ consists of the transformations 
\[
\tilde t=e^{\delta_1}t+\delta_0,\quad
\tilde x=e^{\delta_1}x,\quad
\tilde u=e^{-\frac23\delta_1}u,\quad
\] 
where $\delta_0$ and $\delta_1$ are arbitrary constants. 
At the same time, equation~\eqref{EqFinRemarkable} has remarkable properties connected with different kind of non-Lie symmetries 
that allows us to construct a number of its exact solutions. 
Sources of this singularity should be investigated additionally. 

Due to possibility of changing sign of~$u$ and due to physical sense of the equation, 
we can assume $u$ positive and omit modular in the expression $|u|^{-3/2}$. 

Instead of equation~\eqref{EqFinRemarkable}, we can investigate the equivalent equation
\begin{equation}\label{EqFinRemarkableInV}
v_t=vv_{xx}-\frac23(v_x)^2-\frac32\frac vx,
\end{equation}
where $v=u^{-3/2}$ and, therefore, $v$ is positive. 
In the variables $(t,x,v)$ the operator~$\p_t$ has the same form and $D=t\p_t+x\p_x+v\p_v$.
The inverse transformation from~\eqref{EqFinRemarkableInV} to~\eqref{EqFinRemarkable} is $u=v^{-2/3}$.

\section{List of exact solutions}

For usability we collect all constructed solutions of~\eqref{EqFinRemarkable} together and then discuss 
methods of finding them, including both Lie and non-Lie techniques. 
The adduced solutions are inequivalent with respect to the group~$G_1$ 
and can be extended to parametric sets of solutions of equation~\eqref{EqFinRemarkable} with transformations from this group:
\begin{gather*}
1)\ u=\left(-\varepsilon^2x^3+3\varepsilon x^2-\frac94x\right)^{-2/3}\!, \ \varepsilon\in\{-1,0,1\}\!\!\!\!\mod G_1,
\\
2)\ u=\left(\frac32\frac{x^2}t-\frac94x\right)^{-2/3}\!, 
\\
3)\ u=\left(x^3-3x^2\tan 2t -\frac94x\right)^{-2/3}\!, 
\\
4)\ u=\left(-x^3+3x^2\tanh 2t -\frac94x\right)^{-2/3}\!, 
\\ 
5)\ u=\left(-x^3+3x^2\coth 2t -\frac94x\right)^{-2/3}\!, 
\\
6)\ \sqrt{\psi-\psi^2}-\frac12\arcsin(2\psi-1)=\pm\frac1x+C_0, \quad \psi:=-\frac{u^{-1/2}}{x},\quad 0<\psi<1,
\\
7)\ \sqrt{\psi+\psi^2}-\frac12\ln(2\psi+1+\sqrt{\psi+\psi^2})=\pm\frac1x+C_0, 
\quad \psi:=-\frac{u^{-1/2}}{x},\quad \psi<-1\ \mbox{or}\ \psi>0.
\end{gather*}
Solutions 1)--5) should be considered only for the values of $(t,x)$ where the corresponding bases of power~$-2/3$ is positive.

Note that solutions of~\eqref{EqFinRemarkable} can be transformed to solutions of equations 
which are pointwise equivalent to equation~\eqref{EqFinRemarkable}. 
For example, the transformation $\tilde t=t$, $\tilde x={x}^{-1}$, $\tilde u=x^2u$
links \eqref{EqFinRemarkable} with the equation
\[
{\tilde x}^{-1}\tilde u_{\tilde t}=(\tilde u^{-3/2}\tilde u_{\tilde x})_{\tilde x}+\tilde u.
\]
More generally, the above equation and~\eqref{EqFinRemarkable} belong to the class of variable-coefficient diffusion--reaction equations 
which were investigated in~\cite{Vaneeva&Johnpillai&Popovych&Sophocleous2006}. 
Therefore, equation~\eqref{EqFinRemarkable} can be extended by transformations from the corresponding equivalence group 
to a subclass of variable-coefficient diffusion--reaction equations with simultaneous extension of the above solutions.

\section{Lie reductions}

Let us discuss ways of finding the above solutions starting with the classical Lie method. 
$A_1$ is a non-Abelian two-dimensional Lie algebra. 
A complete list of inequivalent subalgebras of~$A_1$ is exhausted by the one-dimensional subalgebras~$\langle\p_t\rangle$ and $\langle D\rangle$ 
and the algebra~$A_1$ itself. 

With $A_1$ we construct the ansatz $u=\varphi x^{-2/3}$. 
In view of positivity of~$u$, $\varphi$ should be also positive.
Since $A_1$ has a single functionally independent invariant, 
there are no invariant independent variables in the ansatz and $\varphi$ is a 0-ary function i.e. a constant. 
Therefore, the ansatz reduces equation~\eqref{EqFinRemarkable} 
to the algebraic equation~$4\varphi^{-3/2}+9\sign x=0$ with respect to $\varphi$, 
which has a solution only on the negative semiaxis $x<0$.
As a result, we obtain solution 1) with $\varepsilon=0$. 
It is interesting in the sense that all other solutions 1)--5) are modifications of it with additional terms. 
 
The solutions being invariant with respect to the subalgebra~$\langle\p_t\rangle$ are nothing but 
stationary solutions. The corresponding anzatz $u=\varphi(\omega)$, where $\omega=x$, gives the reduced ODE 
\begin{equation}\label{EqFinRemarkableReducedForStationarySolutions}
(\varphi^{-3/2}\varphi_\omega)_\omega+\omega^{-1}\varphi=0
\end{equation}
which is integrable in quadratures. 
Equation~\eqref{EqFinRemarkableReducedForStationarySolutions} is connected with equations 6.101 and 6.205 of~\cite{Kamke1971}.
Integrability of~\eqref{EqFinRemarkableReducedForStationarySolutions} 
can be explained in the framework of symmetry approach. 
The Lie invariance algebra of~\eqref{EqFinRemarkableReducedForStationarySolutions} 
is generated by the operators $\hat D=3\omega\p_\omega-2\varphi\p_\varphi$ and $\hat\Pi=\omega^2\p_\omega-2\omega\varphi\p_\varphi$, 
i.e. it is two-dimensional. 
It is enough for equation~\eqref{EqFinRemarkableReducedForStationarySolutions} to be integrable with the Lie method. 

The invariance algebra $A_1$ induces only the subalgebra spanned the operator~$\hat D$. 
Therefore, $\hat\Pi$ is a pure \emph{hidden} symmetry operator of the initial equation~\eqref{EqFinRemarkable}. 
Let us note that the first non-trivial example of hidden symmetries connected with reduction of PDEs was found 
by Kapitansky~\cite{Kapitanskiy1978,Kapitanskiy1979} for the Navier--Stokes equations. 
Wide classes of hidden symmetries of the Navier--Stokes equations were constructed in~\cite{Fushchych&Popovych1994}.
See also~\cite{Abraham-Shrauner&Leach&Govinder&Ratcliff1995} for different notions of hidden symmetries of ODEs 
and other references therein.

To reduce equation~\eqref{EqFinRemarkableReducedForStationarySolutions} to an integrable form, we change the unknown function as 
\[
\psi=-\frac{\varphi^{-1/2}}\omega, \quad\mbox{i.e.}\quad \varphi=\frac1{\omega^2\psi^2}.
\]
In view of~\eqref{EqFinRemarkableReducedForStationarySolutions}, $\psi_\omega\not=0$ and $\psi$ satisfies the equation 
$(\omega^4\psi_\omega{}^2)_\omega=(\psi^{-1})_\omega$ which is simply integrated once.
Further integration of the obtained first-order ODE $\omega^4\psi_\omega{}^2=\psi^{-1}+C_1$, where $C_1$ is an arbitrary constant, 
with separation of variables results in the implicit solution
\[
\int\frac{d\psi}{\sqrt{\psi^{-1}+C_1}}=\pm\frac1\omega+C_0.
\]  
The integration constant~$C_1$ is normalized to~$\{-1,0,1\}$ by induced scale transformations.
Let us note also that all values of~$C_0$ are equivalent with respect to the hidden symmetry group generated by the operator~$\hat\Pi$.
Calculation of the integral depends on the value of~$C_1$.

If $C_1=0$ then $\psi>0$, i.e. in view of the definition of $\psi$ solutions may exist only for negative values of~$\omega$. 
Integration under the condition $C_1=0$ results in the solution $\psi=(C_0-\frac32x^{-1})^{\frac23}$ 
corresponding to solution~1) of equation~\eqref{EqFinRemarkable}. 
The integration constant~$C_0$ is normalized to~$\varepsilon\in\{-1,0,1\}$ by scale transformations 
associated with the operator~$\hat D$.

The condition $C_1=-1$ implies $0<\psi<1$, i.e.\ solutions again exist only for negative values of~$\omega$ and 
have an implicit form giving solution~6) of equation~\eqref{EqFinRemarkable}. 
In case $C_1=1$ we have the constrain $\psi<-1$ or $\psi>0$ and derive solution~7).

The other kind of Lie invariant solutions are the similarity solutions which are invariant with respect to scale transformations. 
It is more convenient here to work in terms of the variables $(t,x,v)$.
The ansatz constructed with the subalgebra $\langle D\rangle$ has the form $v=t\varphi(\omega)$, where $\omega=x/t$. 
After substituting it to~\eqref{EqFinRemarkableInV}, we obtain the reduced ODE
\[
\varphi\varphi_{\omega\omega}-\frac23(\varphi_\omega)^2+\omega\varphi_\omega-\frac32\frac\varphi\omega-\varphi=0.
\]
It has two polynomial solutions $\varphi=-\frac94\omega$ and $\varphi=\frac32\omega^2-\frac94\omega$ which correspond to 
solutions~$1)_{\varepsilon=0}$ and~2) of equation~\eqref{EqFinRemarkable}.

\section{Non-Lie ansatz}\label{SectionOnNon-LieAnsatz}

The form of Lie invariant solutions~1) and~2) directs us to look for more general polynomial solutions of equation~\eqref{EqFinRemarkableInV}. 
As a result, we find the ansatz 
\[
v=\varphi^1(t)x^3+\varphi^2(t)x^2-\frac94x,
\]
which reduces equation~\eqref{EqFinRemarkableInV} to the system of two ODEs
\[
\varphi^1_t=0, \qquad \varphi^2_t=-6\varphi^1-\frac23(\varphi^2)^2.
\]
Up to translations with respect to~$t$ and scale transformations induced by the Lie symmetry group 
of the initial equation, the above system has the following inequivalent solutions
\[
(\varepsilon^2,\varepsilon), \quad (0,-\tfrac32t), \quad 
(1,-3\tan 2t), \quad (-1,3\tanh 2t), \quad (-1,3\coth 2t)
\]
which correspond to solutions 1)--5) of equation~\eqref{EqFinRemarkable}. 
Solutions 3)--5) are non-Lie ones.

The above ansatz is rewritten in terms of the function~$u$ as 
\[
u=\left(\varphi^1(t)x^3+\varphi^2(t)x^2-\frac94x\right)^{-2/3}.
\]
This ansatz can be interpreted in the framework of a number of different approaches of finding exact solutions 
of nonlinear PDEs, such as
nonlinear variable separation~\cite{Galaktionov1990}, 
the method of differential constraints~\cite{Sidorov&Shapeev&Yanenko}, 
anti-reduction~\cite{Fushchych&Zhdanov1994} or 
generalized conditional symmetries~\cite{Fokas&Liu1994,Zhdanov1995}. 
(See also~\cite{Olver1994} for connections between these approaches.) 
Thus, the differential constraint $2x^3(x^{-2}u^{-3/2})_{xx}=-9$ corresponding to the ansatz 
is compatible (i.e. in involution) with equation~\eqref{EqFinRemarkable}. 
`Anti-reduction' of equation~\eqref{EqFinRemarkable} by the ansatz containing two new unknown functions of one argument 
to the system of two ODEs means that  
\[
(8u^2+8xuu_x+5x^2u_x{}^2-x^2uu_{xx}+6xu^{7/2})\partial_u
\]
is a generalized conditional symmetry operator of equation~\eqref{EqFinRemarkable}.

\section{On nonclassical symmetries}

We also study nonclassical (conditional) symmetries of equation~\eqref{EqFinRemarkable}. 
(See e.g. \cite{Popovych&Vaneeva&Ivanova2005,Zhdanov&Tsyfra&Popovych1999} for necessary definitions and properties 
of nonclassical symmetries.)
Reduction operators of equation~\eqref{EqFinRemarkable} have the general form
$Q=\tau\p_t+\xi\p_x+\eta\p_u$, where $\tau$, $\xi$ and~$\eta$ are functions of $t$, $x$ and $u$, 
and $(\tau,\xi)\ne0$. 
Since \eqref{EqFinRemarkable} is an evolution equation, there are two principally different
cases of finding $Q$: $\tau\ne0$ and $\tau=0$.

We derive the system of determining equations in case of $\tau\ne0$ and integrate it completely. 
As a result, we obtain the following statement. 
\emph{Any conditional symmetry operator of equation~\eqref{EqFinRemarkable} in case of $\tau\ne0$
is equivalent to a Lie symmetry operator.}

As well-known, the operators with the vanishing coefficient of $\partial_t$ form so-called `no-go' case
in study of conditional symmetries of an arbitrary $(1+1)$-dimensional evolution equation
since the problem on their finding is reduced to a single equation which is equivalent to the initial one
(see e.g.~\cite{Fushchych&Shtelen&Serov&Popovych1992,Popovych1998,Zhdanov&Lahno1998}).
Note that ``no-go'' has to be treated as impossibility only of exhaustive solving of the problem. 
A number of particular examples of reduction operators with $\tau=0$ can be constructed  
under additional constraints and then applied to finding exact solutions of the initial equation. 
Since the determining equation has more independent variables and, therefore, more freedom degrees,
it is more convenient often to guess a simple solution or a simple ansatz
for the determining equation, which can give a parametric set of complicated solutions of the initial equation. 
Namely, in the case $\tau=0$ we have $\xi\not=0$. 
Up to usual equivalence of reduction operators, $\xi$ can be assumed equal to 1, i.e. $Q=\p_x+\eta\p_u$. 
The conditional invariance criterion implies the determining equation on
the coefficient~$\eta$
\[
\eta_t=\frac{\eta_{xx}+2\eta\eta_{xu}+\eta^2\eta_{uu}}{u^{3/2}}
-\frac{9\eta\eta_x+6\eta^2\eta_u}{2u^{5/2}}+\frac{15\eta^3}{4u^{7/2}}
+\frac{\eta-u\eta_u}x-\frac{u}{x^2}
\]
which is reduced with a non-point transformation to equation~\eqref{EqFinRemarkable}, 
where $\eta$ becomes a parameter. 
We have found partial solutions of the determining equation but all of them result to the above solutions of 
equation~\eqref{EqFinRemarkable}. In particular, the operator 
\[
\p_x-2\frac ux\Bigl(1\pm\sqrt{C_1u-xu^{3/2}}\,\Bigr)\p_u
\]
gives solutions 1), 6) and 7) in case of the values $C_1=0$, $C_1=-1$ and $C_1=1$ correspondingly.
The ansatz of Section~\ref{SectionOnNon-LieAnsatz} is associated under the condition $\varphi^1=C_1$ with the operator 
\[
\p_x-\frac{u}{6x}\bigl(4C_1x^3u^{\frac 32}+9xu^{\frac 32}+8\bigr)\p_u
\]
which results, therefore, in the solutions 1)--5) depending on values of $C_1$.

\section*{Acknowledgements}
The research of R.\,P. was supported by Austrian Science Fund
(FWF), Lise Meitner project M923-N13. The research of O.\,V. was
partially supported by the grant of the President of Ukraine for
young scientists GF/F11/0061. R.\,P. and O.\,V. are grateful for the hospitality
and financial support by the University of Cyprus.

\end{document}